\documentclass[11pt]{article}
\pdfoutput=1
\usepackage{fullpage}
\usepackage{xcolor}
\usepackage{cite}
\usepackage{graphicx}
\usepackage{tensor}
\usepackage{amsmath}
\usepackage{amssymb}
\usepackage{subfig}

\usepackage[utf8x]{inputenc} 
\title{}
\date{}
\renewcommand{\vec}[1]{\mbox{\boldmath$ #1 $}}

\def\beq{\begin{equation}}
\def\eeq{\end{equation}}
\allowdisplaybreaks
\begin{document}
\bibliographystyle{utphys}

% Commands
\newcommand{\be}{\begin{equation}}
\newcommand{\ee}{\end{equation}}
\newcommand\n[1]{\textcolor{red}{(#1)}} %in-text notes
\newcommand{\diff}{\mathop{}\!\mathrm{d}}
\newcommand{\lb}{\left}
\newcommand{\rb}{\right}
\newcommand{\f}{\frac}
\newcommand{\pd}{\partial}
\newcommand{\tr}{\text{tr}}
\newcommand{\fdiff}{\mathcal{D}}
\newcommand{\im}{\text{im}}
\let\caron\v
\renewcommand{\v}{\mathbf}
\newcommand{\T}{\tensor}
\newcommand{\R}{\mathbb{R}}
\newcommand{\C}{\mathbb{C}}
\newcommand{\Z}{\mathbb{Z}}
\newcommand{\msbar}{\ensuremath{\overline{\text{MS}}}}
\newcommand{\DIS}{\ensuremath{\text{DIS}}}
\newcommand{\abar}{\ensuremath{\bar{\alpha}_S}}
\newcommand{\bb}{\ensuremath{\bar{\beta}_0}}
\newcommand{\rc}{\ensuremath{r_{\text{cut}}}}
\newcommand{\Nd}{\ensuremath{N_{\text{d.o.f.}}}}
\newcommand{\red}[1]{{\color{red} #1}}
%\setlength{\parindent}{0pt}
% Nathan's macros
\newcommand{\mf}[1]{\mathfrak{#1}}
\newcommand{\cl}[1]{\mathcal{#1}}
\renewcommand{\[}{\begin{equation}\begin{aligned}}
\renewcommand{\]}{\end{aligned}\end{equation}}
\titlepage
\begin{flushright}
ADP-24-10/T1249\\
\end{flushright}

\vspace*{0.5cm}

\begin{center}
{\bf \Large The magic of entangled top quarks}

\vspace*{1cm} 
\textsc{Chris D. White$^a$\footnote{christopher.white@qmul.ac.uk} and
Martin J. White$^b$\footnote{martin.white@adelaide.edu.au}} \\

\vspace*{0.5cm} $^a$ Centre for Theoretical Physics, School of Physical and Chemical Sciences, \\
Queen Mary University of London, 327 Mile End
Road, London E1 4NS, UK\\

\vspace*{0.5cm} $^b$ ARC Centre of Excellence for Dark Matter Particle
Physics \& CSSM,  \\ Department of Physics, University of Adelaide,
Adelaide, SA 5005, Australia\\

\end{center}

\vspace*{0.5cm}

\begin{abstract}
Recent years have seen an increasing body of work examining how
quantum entanglement can be measured at high energy particle physics
experiments, thereby complementing traditional table-top
experiments. This raises the question of whether more concepts from
quantum computation can be examined at colliders, and we here consider
the property of {\it magic}, which distinguishes those quantum states
which have a genuine computational advantage over classical states. We
examine top anti-top pair production at the LHC, showing that
nature chooses to produce magic tops, where the amount of magic varies
with the kinematics of the final state. We compare results for individual partonic channels and at proton-level, showing that averaging over final states typically increases magic. This is in contrast to entanglement measures, such as the concurrence, which typically decrease. Our results create new links between the quantum information and particle physics literatures, providing practical insights for further study. 
\end{abstract}

\vspace*{0.5cm}

\section{Introduction}
\label{sec:intro}

It has long been known that there are fundamental limits to the power
of classical computers, so that they are unable to efficiently
simulate the quantum world we live in~\cite{Poplavskii:1975}. This in
turn led to the development of universal quantum computers, replacing
the universal Turing machines of classical computation (see
e.g. ref.~\cite{Nielsen:2012yss} for a pedagogical review). It is an
active field of research to experimentally realise large-scale
quantum computers, and the parallel field of quantum information
theory aims to quantify how information can be encoded, transmitted
and corrected for errors. If quantum computers are to become a
reality, it is imperative that algorithms for quantum computation be
fault tolerant, and thus able to cope with potentially noisy
communication channels.

As well as practical applications, quantum computing and / or
information is also studied in the context of fundamental tests of
quantum mechanics itself, and there is therefore widespread interest
in finding novel physical systems which manifest quantum concepts and
behaviour. To this end, an increasing body of work in recent years has
examined quantum entanglement signatures at particle collider
experiments such as the {\it Large Hadron Collider (LHC)}. Gains for
particle physics include new signatures for disentangling new physics
from the Standard Model. For quantum theorists, collider physics
offers an unprecedentedly high energy regime in which to test their
ideas. In this paper, we will focus on a particular scattering process
at the LHC, namely the production of top-antitop quark pairs. The top
quark is the heaviest known fundamental particle, and thus a
potentially clear window through which to view new physics. It is also
unique among the quarks in that it decays before it has a chance to
hadronise, so that its kinematic properties (such as its spin) can be
efficiently measured. Given that the top quark has spin 1/2, a produced system of a top quark and an antitop quark constitutes a {\it two-qubit system} in the parlance
of quantum information theory. Its use as a probe of quantum
entanglement at the LHC was first proposed in
ref.~\cite{Afik:2020onf,Afik:2022kwm}, resulting in much follow-up
work: see
e.g. refs.~\cite{Aoude:2022imd,Fabbrichesi:2021npl,Severi:2021cnj,Afik:2022kwm,Aoude:2022imd,Aguilar-Saavedra:2022uye,Fabbrichesi:2022ovb,Afik:2022dgh,Severi:2022qjy,Mantani:2022dao,Aguilar-Saavedra:2023hss,Han:2023fci,Simpson:2024hbr,Aguilar-Saavedra:2024hwd,Maltoni:2024csn}
for theoretical studies relating to the top sector alone, and
refs.~\cite{ATLAS:2023jzs,ATLAS:2023fsd,CMS:2024hgo} for experimental
work from both the ATLAS and CMS collaborations. A thorough and
pedagogical recent review of both theory and experiment is that of
ref.~\cite{Barr:2024djo}.

Work to date has typically focused on the entanglement of (anti-)top
spins, and whether this can be convincingly measured. One may also
consider new physics corrections (e.g. expressed in the
model-independent language of effective field
theory~\cite{Aoude:2022imd}), to see whether measured entanglement can
provide clear signatures of theories beyond the Standard Model.
However, and as we will review in detail below, entanglement is not
the only property characterising quantum systems, a point which has
also been made in a top-related context in ref.~\cite{Afik:2022dgh}. A
fruitful interplay between the fields of quantum computing and high
energy physics should involve finding other interesting quantum
information theoretic quantities, and seeing why and how these
manfiest themselves in a collider setting.

In this paper, we will examine a property called {\it magic} which,
roughly speaking, measures whether quantum states possess a
computational advantage over classical states. The study of magic is
pivotal to such questions as how to make fault-tolerant quantum
computers (see e.g.~\cite{Nielsen:2012yss}), and various studies have
addressed how to properly quantify this
concept~\cite{Beverland_2020,PhysRevApplied.19.034052,Leone:2023avk,Qassim2021improvedupperbounds,PhysRevLett.128.050402,Haug:2023ffp,Magic1,PhysRevA.108.042408,Gu:2023och,Tirrito:2023fnw,Turkeshi:2023lqu}. Given
the recent nature of much of this work, it seems particularly timely
to explore whether the property of magic is a natural inevitability at
current collider experiments. Put more simply: does nature produce
magic top quarks and, if not, why not? Such questions are conceptually
interesting in their own right perhaps, but may also clarify how to
exploit magic in other naturally occurring quantum systems. Going the other way, collider analyses provide an experimental playground for developing insights into quantum information theory. Crucial for this is that we can map concepts from
quantum computation to collider physics, and appreciate what they look
like in very practical terms. Our aim, therefore, is to perform a case
study of magic production at colliders, using top pairs in the
Standard Model as an illustrative example. We fully expect our results
to generalise beyond this, and hope that our results prove useful in
the ongoing discussions taking place between two different subfields
of physics.

The structure of our paper is as follows. In section~\ref{sec:review},
we review concepts from quantum computing / information theory,
including the notion of magic. In section~\ref{sec:top}, we classify
in detail how magic is manifested in top pair production at the
LHC. We discuss our results and conclude in section~\ref{sec:discuss}.

\section{Review of relevant concepts}
\label{sec:review}

In this section, we will review salient details of quantum computing
and magic. Given the wider than usual potential audience of this
paper, we aim to be introductory and self-contained, as well as to set
up conventions that will be needed throughout. The basic ingredient of the classical computer is the {\it bit}, that
may assume the binary values 0 and 1. The simplest quantum computers
replace this notion with a {\it qubit}, namely an individual quantum
two-state system whose general state $|\psi\rangle$ is a superposition
of two orthogonal basis states $|0\rangle$ and $|1\rangle$:
\begin{equation}
  |\psi\rangle=\alpha |0\rangle+\beta|1\rangle.
  \label{psidef}
\end{equation}
Here $\alpha$ and $\beta$ are complex numbers satisfying the
normalisation condition
\begin{equation}
  |\alpha|^2+|\beta|^2=1,
  \label{alphabeta}
\end{equation}
and such that the probability of measuring the qubit in the state
$|0\rangle$ or $|1\rangle$ is given by $|\alpha|^2$ or $|\beta|^2$
respectively. A general quantum circuit then takes a number of qubits,
and subjects them to a series of transformations, each of which may
act on single or multiple qubits. In order to conserve probability,
these transformations should be unitary, and each such transformation
is referred to as a {\it quantum gate}, given that this mimics the
action of logic gates on bits in classical computing. A universal
quantum computer is one which has a sufficient number and variety of
quantum gates, such that it can reproduce the effect of an arbitrary
multi-qubit unitary transformation.

There are various ways to represent the effect of quantum
gates. First, let us consider gates that act on a single qubit
individually. The general state of the qubit can be written in 
column form as
\begin{equation}
  |\psi\rangle=\left(\begin{array}{c}\alpha\\ \beta\end{array}\right),
    \label{psicolumn}
\end{equation}
using the basis states appearing in eq.~(\ref{psidef}). The action of
a single-qubit gate will then be a $2\times2$ matrix, and examples
include the {\it phase gate}
\begin{equation}
  P(\phi)=\left(\begin{array}{cc}1&0\\0 & i\end{array}\right)
  \label{phasegate}
\end{equation}
(so-called because it introduces a $\pi/2$ phase shift $\phi$ between the basis
states $|0\rangle$ and $|1\rangle$), and the {\it Hadamard gate}
\begin{equation}
  H=\frac{1}{\sqrt{2}}\left(\begin{array}{cc}1 & 1\\ 1 & -1\end{array}\right),
  \label{Hadamard}
\end{equation}
which is widely used in quantum circuits. For multi-qubit gates, let
us first note that a general $n$-qubit state will be a superposition
of basis states
\begin{equation}
  |\alpha_1\alpha_2\ldots\alpha_n\rangle\equiv
  |\alpha_1\rangle\otimes |\alpha_2\rangle\otimes\ldots\otimes|\alpha_n\rangle,
  \label{nqubit}
\end{equation}
where $\alpha_i\in\{0,1\}$. The number of individual basis states will
be $2^n$, given that each individual qubit has two independent states.
One way to specify the action of a multi-qubit gate is then to choose
some ordering of these basis states, and to represent the action of a
gate as a matrix. As an example, consider the 2-qubit
system with basis states
$(|00\rangle,|01\rangle,|10\rangle,|11\rangle)$. A general 2-qubit
gate will be representable as a $4\times4$ unitary matrix, and a
particular example that is heavily used in quantum computing is the
{\it controlled NOT (CNOT) gate}, whose associated matrix is
\begin{equation}
  {\rm CNOT}=\left(\begin{array}{cccc}
    1 & 0 & 0 & 0\\
    0 & 1 & 0 & 0\\
    0 & 0 & 0 & 1\\
    0 & 0 & 1 & 0
  \end{array}\right).
  \label{CNOT}
\end{equation}
A special class of quantum gates consists of those that act
independently on individual qubits. One may write such gates in tensor
product notation as
\begin{equation}
  U=U_1\otimes U_2\otimes \ldots\otimes U_n,
  \label{Uprod}
\end{equation}
where the action of the gate on a multiqubit state is
\begin{equation}
  U\Big[|\alpha_1\rangle\otimes |\alpha_2\rangle\otimes\ldots\otimes
    |\alpha_n\rangle\Big]=
  \big(U_1|\alpha_1\rangle\big)\otimes
  \big(U_2|\alpha_2\rangle\big)\otimes\ldots\otimes
  \big(U_n|\alpha_n\rangle\big).
  \label{Uprod2}
\end{equation}

Quantum computers differ from their classical counterparts due to the
two key phenomena of {\it superposition} and {\it entanglement}. The
former is already seen in eq.~(\ref{psidef}), and implies that we can
view a multi-qubit quantum computation (before measurement) as a
superposition of a large number of possible classical
computations. Likewise, the creation of entangled states of qubits has
no classical counterpart, and the presence of entanglement in certain
well-known quantum algorithms allows for an exponential speed-up
relative to classical
computing~\cite{365700,doi:10.1137/S0097539795293172}. Despite this,
the na\"{i}ve expectation that entanglement is sufficient to yield
powerful speed-ups over classical computing is not in fact true. For
any number of qubits, there is a certain number of so-called {\it
  stabiliser states}, which include some maximally entangled
configurations, and whose precise definition will be given below. The
{\it Gottesman-Knill theorem}~\cite{Gottesman:1998hu} states that
quantum circuits involving stabiliser states only can be efficiently
simulated using a classical computer. Put more simply, entanglement by
itself is not enough to provide an advantage over classical
algorithms, and something else is therefore needed in order to fully
realise quantum perks. One can then introduce a property that measures
``non-stabiliserness'' of a multi-qubit state, and this property is
known as {\it magic} in the quantum computing literature. Various
definitions have been given to quantify this effect, and the concept
of magic plays a key role in designing fault-tolerant computational
algorithms. Here, we have been inspired by
ref.~\cite{Niroula:2023meg}, which looks at how to increase magic in
larger multi-qubit systems.

Let us now explore the notions of stabiliser circuits and magic in more
detail. Given a (multi-qubit) quantum state $|\psi\rangle$, we say
that it is {\it stabilised} by a unitary operator $U$ provided
\begin{equation}
  U|\psi\rangle=|\psi\rangle.
  \label{stabdef}
\end{equation}
A particular class of quantum states has been well-studied for various
reasons, namely those that are stabilised by elements of the {\it
  Pauli group}. For a single qubit, this group is defined via the set
of matrices
\begin{equation}
  G_1=\{\pm I,\pm iI,\pm \sigma_1,\pm i\sigma_1,\pm \sigma_2,\pm i\sigma_2,
  \pm \sigma_3,\pm i\sigma_3\}.
  \label{G1def}
\end{equation}
Here $I$ is the $2\times 2$ identity matrix, and
$(\sigma_1,\sigma_2,\sigma_3)$ are the Pauli matrices:
\begin{equation}
  \sigma_1=\left(\begin{array}{cc}0 & 1\\1& 0\end{array}\right),\quad
  \sigma_2=\left(\begin{array}{cc}0 & -i\\i& 0\end{array}\right),\quad
    \sigma_3=\left(\begin{array}{cc}1 & 0\\0& -1\end{array}\right).
      \label{XYZdef}
\end{equation}
One may verify that multiplication of any two elements from
eq.~(\ref{G1def}) gives another element of the group, so that this is
indeed closed as required. More generally, one may define the
$n$-qubit Pauli group $G_n$
\begin{equation}
  G_n=\{A_1\otimes A_2\otimes \ldots \otimes A_n; A_i\in G_1\}.
  \label{Gndef}
\end{equation}
In words, the elements of $G_n$ consist of tensor products of
operators $A_i$ from the one-qubit Pauli group, each acting on a
single qubit $i$. Let us define a general {\it Pauli string} as an
operator
\begin{equation}
  {\cal P}_n=P_1\otimes P_2\otimes \ldots\otimes P_N,\quad
  P_a\in\{\sigma^{(a)}_1,\sigma^{(a)}_2,\sigma_3^{(a)},I^{(a)}\},
  \label{Paulistring}
\end{equation}
where e.g. $\sigma^{(a)}$ and $I^{(a)}$ are Pauli and identity matrices
acting on qubit $a$. The elements of the $n$-qubit Pauli group are
then given by Pauli strings weighted by factors $\pm1$, $\pm i$.

The importance of these definitions for quantum computing stems from
the following results (see e.g. Theorem 1 in
ref.~\cite{Aaronson:2004xuh}). Imagine preparing an $n$-qubit quantum
circuit with initial state
\begin{equation}
  |0\rangle^{\otimes n}\equiv |0\rangle\otimes|0\rangle\otimes\ldots\otimes
  |0\rangle.
  \label{0n}
\end{equation}
Now consider that our quantum circuit only contains CNOT, Hadamard and
phase gates. One may then show that $|\psi\rangle$ is stabilised by
exactly $2^n$ Pauli strings (up to factors $\pm1$). Such circuits are
called {\it stabiliser circuits}, and the resulting states are called
{\it stabiliser states}. One may further prove (again see
ref.~\cite{Aaronson:2004xuh}) that the number of independent
stabiliser states for $n$ qubits is
\begin{equation}
  N=2^n\prod_{k=0}^{n-1}\left(2^{n-k}+1\right).
  \label{Nstab}
\end{equation}
As an example, this predicts six stabiliser states for a single qubit
system. These can be found by acting on $|0\rangle$ and $|1\rangle$
repeatedly with Hadamard and phase gates, yielding the six states
\begin{equation}
  |0\rangle, \quad \frac{1}{\sqrt{2}}\left(|0\rangle\pm|1\rangle\right),\quad
  \frac{1}{\sqrt{2}}\left(|0\rangle\pm i|1\rangle\right),\quad |1\rangle.
  \label{stab1}
\end{equation}
From above, we expect each of these states to be an eigenstate of two
operators chosen from $(I,\sigma_1,\sigma_2,\sigma_3)$, where these
two operators could be different for each state. As an example,
$|0\rangle$ is an eigenstate of $I$ and $\sigma_3$ with eigenvalue 1
in both cases (and is not an eigenstate of $\sigma_1$ or
$\sigma_2$). The state $(|0\rangle+i|1\rangle)/\sqrt{2}$ is an
eigenstate of $I$ and $\sigma_2$, with eigenvalue 1. Similar
conclusions can be reached for the other states.

Armed with the concepts of stabiliser circuits and states, we can now
describe the Gottesman-Knill theorem more precisely (see also section
10.5.4 of ref.~\cite{Nielsen:2012yss}). This states that there is an
efficient classical algorithm to simulate any quantum circuit that:
(i) starts with a state prepared in the computational basis; (ii)
contains only Hadamard, phase, CNOT and Pauli gates (i.e. where the
latter implement action of the Pauli matrices on a single qubit);
(iii) involves measurements of observables in the Pauli group; (iv)
may contain additional classical manipulation of the output. In
pedestrian terms, this means that stabiliser circuits provide no
genuine computational advantage over classical circuits, despite the
fact that they may contain maximally entangled states. The presence of
the latter can be clearly seen in eq.~(\ref{stab1}), in which four of
the states are maximally entangled, i.e. consisting of a superposition
of the form of eq.~(\ref{psidef}) with $|\alpha|=|\beta|$.

If superposition and entanglement by themselves are not enough to
guarantee powerful quantum computers, there must be some other
property of quantum states that is relevant. This property is referred
to as {\it magic} in the quantum computing literature, and various
proposals have been given regarding how to quantify
it~\cite{Beverland_2020,PhysRevApplied.19.034052,Leone:2023avk,Qassim2021improvedupperbounds,PhysRevLett.128.050402,Haug:2023ffp,Magic1,PhysRevA.108.042408,Gu:2023och,Tirrito:2023fnw,Turkeshi:2023lqu}. By
definition, stabiliser states have zero magic, and thus one needs to
write down quantities which indeed give zero for stabiliser states,
but give some non-zero measure of ``non-stabiliserness'' for general
states. Here, we will follow the presentation of
ref.~\cite{Turkeshi:2023lqu}, which starts with the above observation
that $n$-qubit stabiliser states are such that the eigenvalues of
$2^n$ Pauli strings are straightforward (i.e. $\pm1$), whereas the others are zero. This suggests that, for a
general state $|\psi\rangle$, we can consider its {\it Pauli spectrum}
\begin{equation}
  {\rm spec}(|\psi\rangle)=\{\langle\psi|P|\psi\rangle,\quad P\in
  {\cal P}_n\}.
  \label{Paulispec}
\end{equation}
That is, the Pauli spectrum of $|\psi\rangle$ is a list of expectation
values of each individual Pauli string in eq.~(\ref{Paulistring}),
evaluated in the state $|\psi\rangle$. For a stabiliser state, this is
a list containing mostly zero values, with $2^n$ non-zero values
$\pm1$. For magic states, a more non-trivial pattern is obtained. From
the Pauli spectrum, one may define something called the {\it
  stabiliser R\'{e}nyi entropies (SREs)}, first defined in
ref.~\cite{Leone:2021rzd}, and consisting of the following family of
functions $\{M_q\}$ defined by integer values $q\geq 2$:
\begin{equation}
  M_q=\frac{1}{1-q}\log_2\left(\zeta_q\right),\quad
  \zeta_q\equiv\sum_{P\in{\cal P}_n}\frac{\langle \psi|P|\psi\rangle^{2q}}
       {2^n}.
       \label{Mqdef}
\end{equation}
We can see that each $M_q$ function involves a sum over expectation
values of the Pauli strings, raised to powers. As argued in
ref.~\cite{Leone:2021rzd} (see also ref.~\cite{Turkeshi:2023lqu}), the
set of values $\{M_q\}$ measures the spread of the Pauli spectrum for
general states, viewed as a distribution. Note that for a stabiliser
state, $M_q=0$ for all $q$. To see this, recall that the expectation
value of $2^n$ of the possible Pauli strings is $\pm 1$ for a
stabiliser state $|\psi\rangle$, and zero for all of the other Pauli strings. Thus,
$\zeta_q=1$ for stabiliser states, and $M_q=\log_2(1)/(1-q)=0$. Below, we
will take the case $q=2$ as a representative measure of magic, for
which $M_2$ is known as the {\it Second Stabiliser R\'{e}nyi entropy
  (SSRE)}.

\section{Are top quarks magic?}
\label{sec:top}

\subsection{Results for partonic initial states}

\begin{figure}
  \begin{center}
    \scalebox{0.9}{\includegraphics{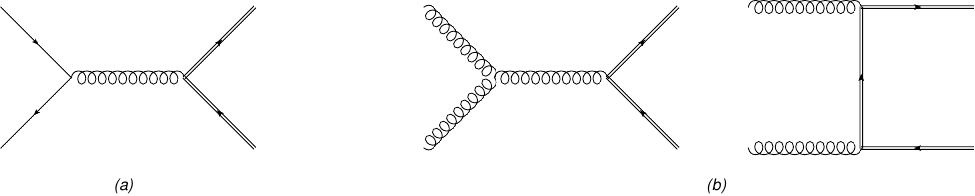}}
    \caption{LO Feynman diagrams for top-antitop pair production in
      the SM, where a double line represents a top particle: (a)
      $q\bar{q}$ channel; (b) $gg$ channel. The two channels
      contribute roughly $10\%$ and $90\%$ of the total cross-section
      respectively.}
    \label{fig:topLO}
  \end{center}
\end{figure}

Let us now turn to the question of whether magic can be probed in
collider physics. To this end, we will consider a well-studied system
from the point of view of entanglement, namely the production of a
top-antitop pair at the LHC. Given that top quarks are spin-1/2
particles, this constitutes a narturally produced two-qubit system. At
leading order in the Standard Model, top quark pair production
proceeds via the strong interaction, and there are two partonic
channels as shown in figure~\ref{fig:topLO}. In each case, QFT
predicts the possible states of the top quarks (e.g. their momenta,
spins and energies). The top quarks will not be in a uniquely
predicted state, but instead must be regarded as in a statistical
mixture of possible states, with associated probabilities. There is a
well-developed formalism for dealing with such situations in quantum
theories, known as the {\it density matrix formalism}. To introduce
this in general, consider a quantum system that can be in any one of a
number of states $|\psi_i\rangle$, each with probability $p_i$. One
may then define the {\it density matrix}
\begin{equation}
  \rho=\sum_i p_i |\psi_i\rangle\langle\psi_i|,
  \label{rhodef}
\end{equation}
where the probabilities must necessarily satisfy
\begin{equation}
  p_i\geq 0,\quad \sum_i p_i=1.
\label{probs}
\end{equation}
If a given system is found to be in a single overall state
$|\psi\rangle$ (with unit probability), its density matrix has the
special form
\begin{equation}
  \rho=|\psi\rangle\langle\psi|,
  \label{rhopure}
\end{equation}
and the system is said to be in a {\it pure state}. If instead two or
more probabilities in eq.~(\ref{rhodef}) are non-zero, then the system
is said to be in a {\it mixed state}. As stated above, it is this
latter case that will apply to top quarks produced by a collider
experiment, as the rules of QFT give us at most the identity of each
possible final state, and its associated probability. Faced with such
an ensemble of possible final states, the expectation value of an
observable ${\cal O}$ will be given by
\begin{align}
  \langle O\rangle=\sum_i p_i \langle \psi_i|{\cal O}|\psi_i\rangle
  ={\rm Tr}[\rho {\cal O}],
  \label{Oexp}
\end{align}
where the second line recognises that this is the trace of the density
matrix multiplied by the operator in matrix form. 

The language of density matrices is used heavily in the context of
entanglement, due to the ease of defining clear criteria for states to
be entangled. Considering two subsystems $A$ and $B$ of a larger
whole, the density matrix $\rho$ is {\it separable} if it can be
written in the form
\begin{equation}
  \rho=\sum_i p_i \rho_i^A\otimes \rho_i^B.
  \label{separable}
\end{equation}
Separable states contain no quantum entanglement between the
subsystems $A$ and $B$. Similarly, entangled states cannot be written
in a separable form. This in turn allows various algebraic criteria
for entanglement to be obtained (see
e.g. refs.~\cite{Afik:2020onf,Afik:2022kwm,Barr:2024djo} for a
review), which in turn have guided experimental searches for top quark
entanglement. In such analyses, it is entanglement of the (anti-)top
quark spins that is being talked about, and there will be a density
matrix describing the (in general mixed) state of a top-antitop
pair. The definition of this can be found in
e.g. ref.~\cite{Afik:2022kwm}, and there are different levels at which
one can define it. First, let us note that the kinematics of a pair of
top particles in the centre of mass frame will be specified by their
invariant mass $M$, the unit direction vector $\vec{k}$ of one of the
particles, and the spins $\alpha$ and $\beta$ of each particle. These
quantities can be used to specify a complete set of possible final
states $|M\hat{\vec{k}}\alpha\beta\rangle$. If we then let
$|I\lambda\rangle$ denote a given initial state $I$ (with associated
quantum numbers $\lambda$), then the quantum (scattering) amplitude
for proceeding from this initial state to one of the possible final
states is, by definition,
\begin{equation}
  {\cal A}^\lambda_{\alpha\beta}=\langle M\hat{\vec{k}}\alpha\beta|\hat{T}|I\lambda\rangle
  \label{ampdef}
\end{equation}
where $\hat{T}$ is the {\it transition operator}. We can then define
the so-called $R$-matrix
\begin{equation}
  R^{I\lambda}_{\alpha\beta,\alpha'\beta'}(M,\hat{\vec{k}})=
  {\cal A}^\lambda_{\alpha\beta}(M,\hat{\vec{k}})
  \left[{\cal A}^\lambda_{\alpha'\beta'}(M,\hat{\vec{k}})\right]^\dag.
  \label{Rmatrix}
\end{equation}
          
This is like a squared amplitude that enters a cross-section, except
for the fact that the final states are different, and thus the spin
indices will not be contracted. In a collider experiment, we do not
usually have control over the precise initial states, which must
therefore be averaged over initial quantum numbers (e.g. colours,
spins etc.). Thus, we can define the averaged $R$-matrix:
\begin{equation}
  R^I_{\alpha\beta,\alpha'\beta'}(M,\hat{\vec{k}})=\frac{1}{N_\lambda}
  \sum_{\lambda} R^{I\lambda}_{\alpha\beta,\alpha'\beta'}(M,\hat{\vec{k}}).
  \label{Rmatrix2}
\end{equation}
This is sometimes loosely referred to as the {\it production (spin)
  density matrix}, but strictly speaking this should be normalised to
have unit trace:
\begin{equation}
  \rho^I=\frac{R^I}{{\rm Tr}(R^I)}.
  \label{rhoIdef}
\end{equation}
This is not the only density matrix we can talk about to describe the
top quark final state. We might also wish to describe the top quark
mixed state that results after we sum over all possible final states
with a given invariant mass (i.e. summing over all possible top quark
directions). The resulting density matrix is given
by~\cite{Afik:2022kwm}
\begin{equation}
  \bar{\rho}^I(M)=\frac{1}{Z}\sum_{\alpha\beta,\alpha'\beta'}\int d\Omega\,
  R^I_{\alpha\beta,\alpha'\beta'}(M,\hat{\vec{k}})\frac{|M\hat{\vec{k}}\alpha\beta\rangle
    \langle M\hat{\vec{k}}\alpha'\beta'|}{\langle M\hat{\vec{k}}|M\hat{\vec{k}}\rangle}.
  \label{rhobarI}
\end{equation}
Here, the denominator of the integrand normalises the final states,
which are then weighted with the production density matrix. There is
then an integral over all possible solid angles (i.e. over the top
quark directions), and the overall prefactor contains a normalisation
factor $Z$. 

\begin{figure}
  \begin{center}
    \scalebox{0.6}{\includegraphics{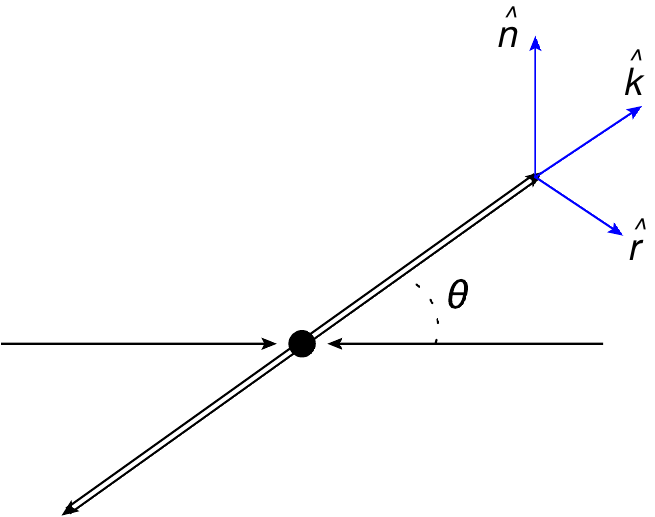}}
    \caption{Illustration of the helicity basis, used to define the
      coefficients of the top quark production density matrix. The
      single and double arrows represent the incoming beams and
      outgoing top quark directions respectively, with $\theta$ the
      scattering angle. The top quark direction is given by the unit
      vector $\hat{\vec{k}}$, where $\hat{\vec{n}}$ and $\vec{r}$ are
      defined in the plane transverse to this.}
    \label{fig:helicity}
  \end{center}
\end{figure}

Let us focus on the production density matrix $R^I$, which is fully
differential in the final state kinematics. This quantity specifies a
general mixed state of the top-antitop pair, and will depend upon the
invariant mass of the latter, and the directions of the top particles
in general. This is what we expect of course: QFT tells us that the
probability of producing a given final state depends upon its
kinematics. The ``mixed'' top quark final state constitutes a
superposition of the possible final states for each point in the kinematic
phase space, weighted by the appropriate probabilities as predicted by
QFT. Results for the top quark production density matrix, for
the initial states $I=q\bar{q}$, $gg$, can be found e.g. in
ref.~\cite{Afik:2022kwm}. The first step in presenting this is to note
that a two-qubit $R$-matrix has the general decomposition
\begin{equation}
  R^I=\tilde{A}^I I_4+\sum_i\left(\tilde{B}^{I+}_i \sigma_i\otimes I_2
  +\tilde{B}^{I-}_i I_2\otimes\sigma_i+\sum_{i,j}\tilde{C}^I_{ij}\sigma_i
  \otimes\sigma_j\right).
  \label{Rdecomp}
\end{equation}
Here $I_n$ denotes an $n$-dimensional identity matrix, and $\sigma_i$
a Pauli matrix. The kinematic dependence of the $R$-matrix is then
sitting in the various coefficients
$\{\tilde{A}^I,\tilde{B}^{I\pm}_i,\tilde{C}^I_{ij}\}$. Upon taking the
trace of eq.~(\ref{Rdecomp}), only the first term survives, such that
the normalised production density matrix is given by
\begin{equation}
  \rho^I=\frac{R^I}{4\tilde{A}^I}.
  \label{rhoIdecomp}
\end{equation}
We can then use a similar decomposition to eq.~(\ref{Rdecomp}) for the
normalised density matrix, with coefficients
\begin{equation}
  B^{I\pm}_i=\frac{\tilde{B}^{I\pm}_i}{\tilde{A}^I},\quad C^I_{ij}=
  \frac{\tilde{C}^I_{ij}}{\tilde{A}^I}.
  \label{BCnorm}
\end{equation}
To give explicit forms for the coefficients, we need to choose a
convenient coordinate system, and a common choice is the so-called
{\it helicity basis}~\cite{Baumgart:2012ay}, illustrated in
figure~\ref{fig:helicity}. This introduces two vectors $\hat{\vec{n}}$
and $\hat{\vec{r}}$ transverse to the top quark direction, which can
be formally defined as follows:
\begin{equation}
  \hat{r}=\frac{\hat{\vec{p}}-\cos\theta\,\hat{\vec{k}}}{\sin(\theta)},\quad
  \hat{\vec{n}}=\hat{\vec{r}}\times\hat{\vec{k}},
  \label{rndef}
\end{equation}
where $\hat{\vec{p}}$ is the direction of the first incoming beam, and
$\cos\theta=\hat{\vec{k}}\cdot\hat{\vec{p}}$ defines the scattering
angle. As noted in ref.~\cite{Afik:2022kwm}, a particular advantage of
the helicity basis is that a boost along the top quark direction does
not change its orientation. Thus, spin information defined in the rest
frame of each top particle is easily relatable to the centre of mass
frame in which the top quarks are measured. One may furthermore show
that the following relations hold at LO in the SM:
\begin{equation}
  \tilde{B}_i^{I+}=\tilde{B}_i^{I-}=0,\quad \tilde{C}^I_{ij}=\tilde{C}^I_{ji},
  \label{BCrels}
\end{equation}
also that certain coefficients in the helicity basis vanish:
\begin{equation}
  \tilde{C}^I_{nr}=\tilde{C}^I_{nk}=0.
  \label{BCrels2}
\end{equation}
The remaining non-zero coefficients will be functions of the top quark
invariant mass and scattering angle. Rather than the former, it is
conventional to introduce the variable
\begin{equation}
  \beta=\sqrt{1-\frac{4m_t^2}{M}},
    \label{betadef}
\end{equation}
which can be shown to be the velocity of the top quark (in natural
units), with $m_t$ the top quark mass. We can see that this makes
sense from eq.~(\ref{betadef}): $\beta=0$ or 1 at threshold ($M=2m_t$)
or high energy respectively, where the top quark velocity in the
latter case approaches the speed of light. Defining also
\begin{equation}
  z=\cos\theta,
  \label{zdef}
\end{equation}
we may quote for the non-zero $R$-matrix coefficients at LO in the SM
from ref.~\cite{Aoude:2022imd}. For the $q\bar{q}$ and $gg$ channels
respectively one has
\begin{align}
  \tilde{A}^{q\bar{q}}&=F_{q\bar{q}}\Big(2+\beta^2(z^2-1)\Big);\notag\\
  \tilde{C}_{nn}^{q\bar{q}}&=F_{q\bar{q}}\beta^2(z^2-1);\notag\\
  \tilde{C}_{kk}^{q\bar{q}}&=F_{q\bar{q}}\Big(\beta^2+z^2(2-\beta^2)\Big);
  \notag\\
  \tilde{C}_{rr}^{q\bar{q}}&=F_{q\bar{q}}\Big(2-\beta^2-z^2(2-\beta^2)\Big);
  \notag\\
  \tilde{C}_{rk}^{q\bar{q}}&=2zF_{q\bar{q}}\sqrt{(1-z^2)(1-\beta^2)};
\label{coeffsqq}
\end{align}
and
\begin{align}
  \tilde{A}^{gg}&=F_{gg}\Big(
  1+2\beta^2(1-z^2)-\beta^4(z^4-2z^2+2)\Big);\notag\\
  \tilde{C}_{nn}^{gg}&=-F_{gg}\Big(1-2\beta^2+\beta^4(z^4-2z^2+2)\Big);\notag\\
  \tilde{C}^{gg}_{kk}&=-F_{gg}\Big(1-2z^2(1-z^2)\beta^2-(2-2z^2+z^4)\beta^4
  \Big);\notag\\
  \tilde{C}^{gg}_{rr}&=-F_{gg}\Big(1-(2-2z^2+z^4)\beta^2(2-\beta^2)\Big);
  \notag\\
  \tilde{C}^{gg}_{rk}&=2zF_{gg} (1-z^2)^{3/2}\beta^2\sqrt{1-\beta^2};
  \label{coeffsgg}
\end{align}
where 
\begin{equation}
  F_{q\bar{q}}=\frac{1}{18},\quad
  F_{gg}=\frac{7+9\beta^2z^2}{192(1-\beta^2z^2)^2}.
  \label{Fggdef}
\end{equation}

\begin{figure}
    \centering
    \subfloat[]{\includegraphics[width=0.45\textwidth]{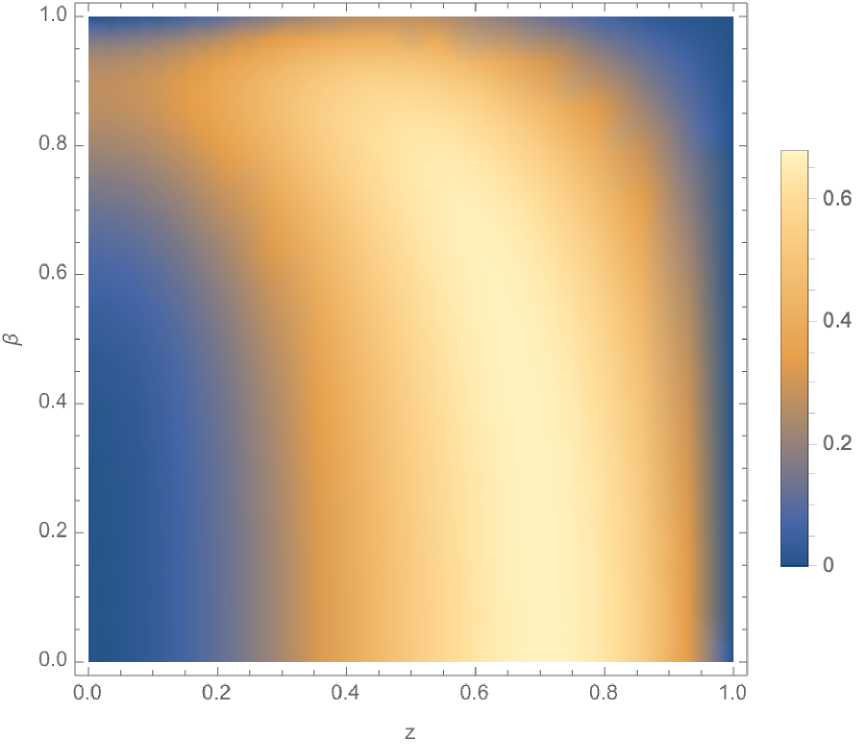} }
    \qquad
    \subfloat[]{\includegraphics[width=0.45\textwidth]{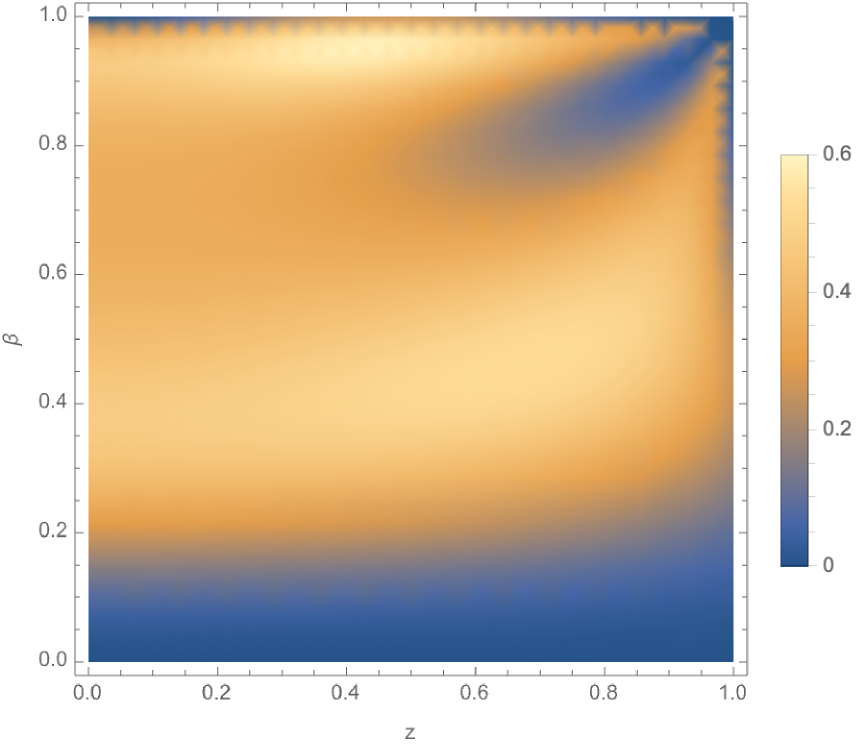} }
    \caption{The magic of a mixed top-antitop final state in: (a) the
      $q\bar{q}$ channel; (b) the $gg$ channel.}
    \label{fig:magic}
\end{figure}

Using these results, one may evaluate the magic of a top-antitop pair,
for different partonic channels and / or kinematics. To do so, we must
first note that the definition of the SSRE $M_2$ given in
eq.~(\ref{Mqdef}) is for pure states, whereas we are instead dealing
with a mixed state. One may, however, generalise the above notions to
this context. First, as discussed in ref.~\cite{Leone:2021rzd} (see in
particular the supplementary material), an $n$-qubit mixed stabiliser
state is defined by having a density matrix of the form
\begin{equation}
  \rho=\frac{1}{2^n}\left(
  I_4+\sum_{P\in G}\phi_P P
  \right).
  \label{rhomixed}
\end{equation}
Here $P$ is a Pauli string; $G$ a subset of the $n$-qubit Pauli group
with $0<|G|<d-1$ elements; and $\phi_P=\pm 1$ for each $P$. Such a
state has a Pauli spectrum containing trivial values $\pm1$, and thus
is a natural generalisation of the concept of a pure stabiliser
state. It turns out also that there is a simple tweak of the
definition of $M_2$ such that this is zero for mixed stabiliser
states. First, let us note that the pure state definition of $M_q$ of
eq.~(\ref{Mqdef}) contains the quantity
\begin{equation}
  \langle \psi|P|\psi\rangle={\rm Tr}(P\rho),
  \label{Trid}
\end{equation}
where $P$ is a Pauli string, and $\rho$ the density matrix
corresponding to the (pure) state $|\psi\rangle$. The formula for
$M_2$ can then be written as
\begin{equation}
  M_2(\rho)=-\log_2\left(\frac{\sum_{P\in{\cal P}_n}{\rm Tr}^4(\rho P)}
  {2^n}\right).
  \label{M2def}
\end{equation}
The appropriate modification of this for mixed states
is~\cite{Leone:2021rzd}
\begin{equation}
  \tilde{M}_2(\rho)=-\log_2\left(\frac{\sum_{P\in{\cal P}_n}{\rm Tr}^4(\rho P)}
  {\sum_{P\in{\cal P}_n}{\rm Tr}^2(\rho P)}\right),
  \label{M2tildedef}
\end{equation}
and plugging in the decomposition of eqs.~(\ref{Rdecomp},
\ref{rhoIdecomp}) eventually gives
\begin{equation}
  \tilde{M}_2(\rho^I)=
  -\log_2\left(\frac{(\tilde{A}^I)^4+(\tilde{C}_{nn}^I)^4
    +(\tilde{C}_{kk}^I)^4+(\tilde{C}_{rr}^I)^4+2(\tilde{C}_{rk}^I)^4}
  {(\tilde{A}^I)^2[(\tilde{A}^I)^2+(\tilde{C}_{nn}^I)^2
    +(\tilde{C}_{kk}^I)^2+(\tilde{C}_{rr}^I)^2+2(\tilde{C}_{rk}^I)^2]}
  \right).
  \label{M2tilderes}
\end{equation}
Using eqs.~(\ref{coeffsqq}, \ref{coeffsgg}), we can now plot the
amount of magic in a top-antitop pair, as measured by $\tilde{M}_2$,
and as predicted by the SM at LO. Results are shown in the $(z,\beta)$
plane in figure~\ref{fig:magic}, where we have used a top quark mass
of $m_t=172.76$GeV. There are a number of notable features. First, it
is relatively generic that magic top quark states are produced
throughout most of phase space. There are then several regions or
limits in which the magic vanishes completely. This can be related to
known results in the literature~\cite{Afik:2022kwm}, as we discuss in
the following section.

\subsection{When has the magic gone?}
\label{sec:vanish}

In order to analyse where the SSRE is zero, it will be useful to find
the form of the $R$-matrix in different kinematic limits. Following
ref.~\cite{Afik:2022kwm}, one may then relate it to specific quantum
states, which gives further physical insight into why magic
vanishes. We can write each of the operators in eq.~(\ref{Rdecomp}) in
terms of an explicit $4\times4$ matrix acting on the computational
basis $(|00\rangle,|01\rangle,|10\rangle,|11\rangle)$, where for each
particle $|0\rangle$ and $|1\rangle$ represent a spin up or down state
in the $\vec{k}$ direction respectively. In terms of the coordinate
basis of figure~\ref{fig:helicity}, this means that we define Pauli
matrices
\begin{equation}
  \sigma_n=\sigma_1,\quad \sigma_r=\sigma_2,\quad \sigma_k=\sigma_3.
  \label{sigmamap}
\end{equation}
In other words, the (1,2,3) directions in spin space (represented by
the Pauli matrices of eq.~(\ref{XYZdef})) are mapped to the
$(\vec{n},\vec{r},\vec{k})$ directions respectively, which then
constitute a right-handed coordinate system. Consider then the example
operator $\sigma_n\otimes\sigma_n$. An explicit computation gives
\begin{align}
  \sigma_n\otimes\sigma_n |00\rangle&=
  \left[\sigma_n |0\rangle\right]\otimes
  \left[\sigma_n |0\rangle\right]\notag\\
  &=\left[
    \left(\begin{array}{cc}0 & 1 \\ 1 & 0\end{array}\right)
      \left(\begin{array}{c}1 \\ 0\end{array}\right)
        \right]\otimes
  \left[
    \left(\begin{array}{cc}0 & 1 \\ 1 & 0\end{array}\right)
      \left(\begin{array}{c}1 \\ 0\end{array}\right)
        \right]\notag\\
  &=\left[
      \left(\begin{array}{c}0 \\ 1\end{array}\right)
        \right]\otimes
  \left[
      \left(\begin{array}{c}0 \\ 1\end{array}\right)
        \right]\notag\\
  &=|11\rangle.
  \label{nnact}
\end{align}
Likewise, we find
\begin{equation}
  \sigma_n\otimes\sigma_n|01\rangle=|10\rangle,\quad
  \sigma_n\otimes\sigma_n|10\rangle=|01\rangle,\quad
  \sigma_n\otimes\sigma_n|11\rangle=|00\rangle,
\label{nnact2}
\end{equation}
such that the matrix form of the operator in the computational basis is
\begin{equation}
  \sigma_n\otimes\sigma_n=\left(\begin{array}{cccc}
    0 & 0 & 0 & 1\\
    0 & 0 & 1 & 0\\
    0 & 1 & 0 & 0\\
    1 & 0 & 0 & 0    
  \end{array}\right).
  \label{sigmanmat}
\end{equation}
Using similar methods, one finds the following:
\begin{align}
  \sigma_k\otimes\sigma_k&=\left(\begin{array}{cccc}
    1 & 0 & 0 & 0\\
    0 & -1 & 0 & 0\\
    0 & 0 & -1 & 0\\
    0 & 0 & 0 & 1    
  \end{array}\right),\quad
  \sigma_r\otimes\sigma_r=\left(\begin{array}{cccc}
    0 & 0 & 0 & -1\\
    0 & 0 & 1 & 0\\
    0 & 1 & 0 & 0\\
    -1 & 0 & 0 & 0    
  \end{array}\right),\notag\\
  \sigma_r\otimes\sigma_k&=\left(\begin{array}{cccc}
    0 & 0 & i & 0\\
    0 & 0 & 0 & -i\\
    -i & 0 & 0 & 0\\
    0 & i & 0 & 0    
  \end{array}\right),\quad
  \sigma_k\otimes\sigma_r=\left(\begin{array}{cccc}
    0 & i & 0 & 0\\
    -i & 0 & 0 & 0\\
    0 & 0 & 0 & -i\\
    0 & 0 & i & 0    
  \end{array}\right).
  \label{sigmamats}
\end{align}
Due to the fact that only certain coefficients in eq.~(\ref{Rdecomp})
are non-vanishing at LO in the SM, one has the simplified
decomposition
\begin{align}
  R^I(z,\beta)&=\tilde{A}^I I_4+\tilde{C}^I_{nn}\sigma_n\otimes\sigma_n
  +\tilde{C}^I_{rr}\sigma_r\otimes\sigma_r
  +\tilde{C}^I_{kk}\sigma_k\otimes\sigma_k
  +\tilde{C}^I_{rk}\left(\sigma_r\otimes\sigma_k+\sigma_k\otimes
  \sigma_r\right),
  \label{Rdecomp2}
\end{align}
where we have used the symmetry property of the $\tilde{C}^I_{ij}$
coefficients noted in eq.~(\ref{BCrels2}). By plugging in the results
of eqs.~(\ref{coeffsgg}, \ref{coeffsqq}), one then has a complete form
for the $R$-matrix in the computational basis, with full kinematic
dependence.

Now let us look at the regions in figure~\ref{fig:magic} where the
magic vanishes, and see what happens to the $R$-matrix in each
case. For the $q\bar{q}$ channel, these limits give
\begin{equation}
  R^{q\bar{q}}(1,\beta)\propto\left(\begin{array}{cccc}
    1 & 0  & 0  & 0\\ 0 & 0 & 0 & 0\\ 0 & 0 & 0 & 0\\ 0 & 0 & 0 & 1
  \end{array}\right),\quad
  R^{q\bar{q}}(0,1)\propto\left(\begin{array}{cccc}
    1 &0 &0 & -1\\ 0 & 0 & 0 & 0\\ 0 & 0 & 0 & 0\\ -1 & 0 & 0 & 1
  \end{array}\right),\quad
  R^{q\bar{q}}(0,0)\propto\left(\begin{array}{cccc}
    1 &0 &0 & -1\\ 0 & 1 & 1 & 0\\ 0 & 1 & 1 & 0\\ -1 & 0 & 0 & 1
  \end{array}\right).
  \label{Rqqlims}
\end{equation}
One may show that each of these matrices indeed meets the definition
of stabiliserness given in eq.~(\ref{rhomixed}), in that each matrix
can be written as a suitable linear combination of Pauli strings, with
coefficients normalised appropriately. To gain further insight into
what is happening to the top quarks in each limit, we may start with
the first matrix in eq.~(\ref{Rqqlims}), and note that this is equal
to
\begin{displaymath}
  \left(\begin{array}{c}
    1\\0\\0\\0
  \end{array}\right)
    \left(\begin{array}{cccc}
    1&0&0&0
    \end{array}\right)+
  \left(\begin{array}{c}
    0\\0\\0\\1
  \end{array}\right)
    \left(\begin{array}{cccc}
    0&0&0&1
    \end{array}\right),
\end{displaymath}
which when translated into state notation yields
\begin{equation}
  R^{q\bar{q}}(1,\beta)\propto
  |00\rangle\langle 00|+|11\rangle\langle 11|.
  \label{Rlim2}
\end{equation}
The density matrix is given by normalising this to have unit trace:
\begin{equation}
  \rho^{q\bar{q}}(1,\beta)=\frac{|00\rangle\langle 00|+|11\rangle\langle 11|}
      {2}.
      \label{Rlim3}
\end{equation}
This agrees with eq.~(26) of ref.~\cite{Afik:2020onf}, where that
reference quotes the result as
\begin{displaymath}
  \frac{|\uparrow_{\hat{p}}\uparrow_{\hat{p}}\rangle
    \langle \uparrow_{\hat{p}}\uparrow_{\hat{p}}|
      +|\downarrow_{\hat{p}}\,\downarrow_{\hat{p}}\rangle
      \langle \downarrow_{\hat{p}}\,\downarrow_{\hat{p}}|}
  {2},
\end{displaymath}
where $|\uparrow_{\hat{p}}\rangle$ and $|\downarrow_{\hat{p}}\rangle$
denote spin up and down states along the beam direction. That this is
equivalent to eq.~(\ref{Rlim3}) follows from the fact that, if the
scattering angle is 0$^\circ$ so that $z=\cos(\theta)=1$, then the
top quark direction $\hat{\vec{k}}$ is the same as the beam direction
$\hat{\vec{p}}$.

It is also possible to describe the other $R$-matrices in
eq.~(\ref{Rqqlims}) in terms of specific states. For the second, and
inspired by refs.~\cite{Afik:2020onf,Aoude:2022imd}, let us define the
single qubit states
\begin{equation}
  |\uparrow_{\hat{n}}\rangle=\frac{|0\rangle+|1\rangle}{\sqrt{2}},\quad
  |\downarrow_{\hat{n}}\rangle=\frac{|0\rangle-|1\rangle}{\sqrt{2}},\quad
  \label{spinn}
\end{equation}
which constitute spin up and down states in the $\hat{\vec{n}}$
direction respectively\footnote{To see this, note that $(1,1)$ and
$(1,-1)$ are eigenstates of the Pauli matrix $\sigma_n$ with
eigenvalues $\pm1$.}. Then, we can define the two-qubit states
\begin{equation}
  |\Psi^\pm\rangle=\frac{|\uparrow_{\hat{n}}\downarrow_{\hat{n}}\rangle
    \pm|\downarrow_{\hat{n}}\uparrow_{\hat{n}}\rangle}{\sqrt{2}},
\label{Psipmdef}
\end{equation}
which substitution of eq.~(\ref{spinn}) reveals to be given in the
computational basis by
\begin{equation}
  |\Psi^+\rangle=\frac{|00\rangle-|11\rangle}{\sqrt{2}}\equiv
  \frac{1}{\sqrt{2}}\left(\begin{array}{c}1\\0\\0\\-1
  \end{array}\right),\quad
  |\Psi^-\rangle=\frac{|10\rangle-|01\rangle}{\sqrt{2}}
  \equiv
  \frac{1}{\sqrt{2}}\left(\begin{array}{c}0\\-1\\1\\0
  \end{array}\right).
  \label{Psi+-res}
\end{equation}
The column vector form can be used to show that the normalised density
matrix corresponding to the second $R$-matrix in eq.~(\ref{Rqqlims})
is given by
\begin{equation}
  \rho^{q\bar{q}}(0,1)=|\Psi^+\rangle\langle\Psi^+|,
  \label{rhoqq01}
\end{equation}
which agrees with eq.~(24) of ref.~\cite{Afik:2020onf}. Using similar
techniques, the final matrix in eq.~(\ref{Rqqlims}) can be shown to
give rise to the density matrix
\begin{equation}
  \rho^{q\bar{q}}(0,0)=\frac{|\Psi^+\rangle\langle\Psi^+|+|\Psi_1\rangle
    \langle\Psi_1|}{\sqrt{2}},\quad |\Psi_1\rangle=\frac{|01\rangle+
  |10\rangle}{\sqrt{2}}.
  \label{Rq00}
\end{equation}
From eq.~(\ref{rhoqq01}), we see that $\rho^{q\bar{q}}(0,1)$ describes
a pure state. Furthermore, this is maximally entangled, as can be seen
from the definition of $|\Psi^+\rangle$ in eq.~(\ref{Psipmdef}). It is
in fact a known two-qubit stabiliser state, and corroborates the fact
that maximal entanglement can be associated with vanishing magic. By
contrast, eqs.~(\ref{Rlim3}, \ref{Rq00}) constitute so-called {\it
  separable mixed states}, for which no quantum entanglement is
present. Again the magic vanishes, so that the kinematic region of
non-zero magic lies in between the two extremes of minimal and maximal
entanglement.

Similar conclusions can be reached for the $gg$ channel. In that case,
the kinematic limits with vanishing magic, and their associated
quantum states, are given by
\begin{equation}
  \rho^{gg}(1,1)=\frac{|00\rangle\langle 00|+|11\rangle\langle 11|}
      {2},\quad \rho^{gg}(z,0)=\frac{|\uparrow_{\hat{n}}\downarrow_{\hat{n}}
        \rangle
        -|\downarrow_{\hat{n}}\uparrow_{\hat{n}}\rangle}{\sqrt{2}},\quad
      \rho^{gg}(0,1)=|\Psi^+\rangle\langle\Psi^+|.
      \label{rhogglims}
\end{equation}
Again these are minimally and maximally entangled, with
the region of non-zero magic lying in between. 

\subsection{Results for proton initial states}
\label{sec:hadron}

\begin{figure}
  \begin{center}
    \scalebox{0.5}{\includegraphics{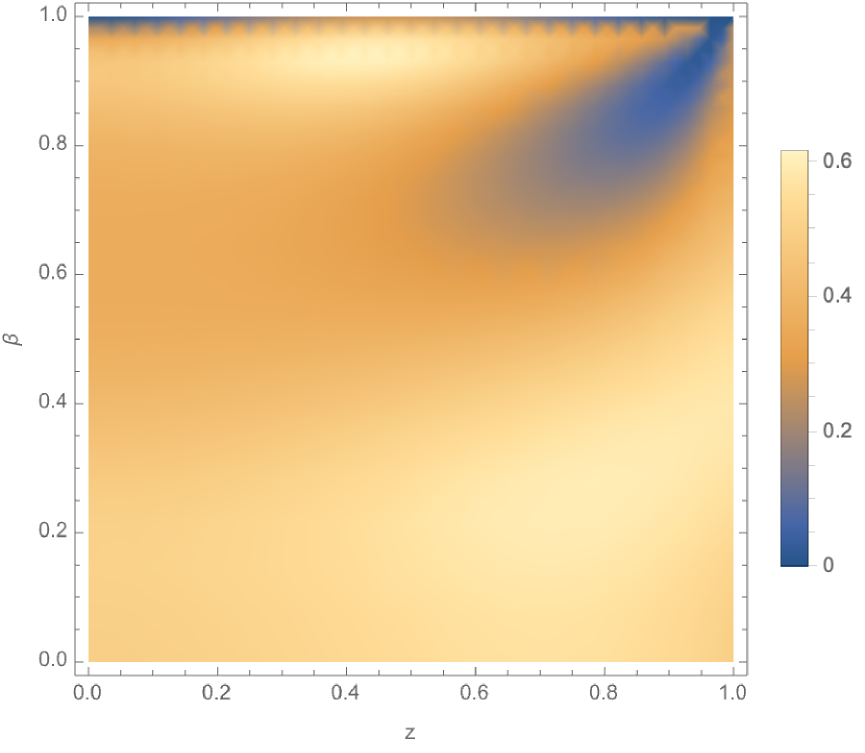}}
    \caption{The magic of a mixed top-antitop final state at hadron
      level.}
    \label{fig:magicpp}
  \end{center}
\end{figure}

Above, we have presented results for individual partonic
channels. However, the total top quark state produced at the LHC will
be a weighted sum of partonic channels, taking into account the
non-perturbative distributions of the quarks and gluons inside the
colliding protons. As stated in ref.~\cite{Afik:2020onf}, it is
straightforward to convert the individual partonic $R$-matrices of
eq.~(\ref{Rmatrix}) to their hadronic counterparts by forming the
combination
\begin{equation}
  R(M,\hat{\vec{k}})=\sum_{I\in\{q\bar{q},gg\}}L^{I}(M)R^I(M,\hat{\vec{k}}),
  \label{Rhad}
\end{equation}
where we have introduced the parton luminosities
\begin{equation}
  L^{I}(M)=\frac{1}{s}\frac{1}{1+\delta_{ab}}\int_\tau^1\frac{dx}{x}f_a(x,\sqrt{\hat{s}})
  f_b\left(\frac{\tau}{x},\sqrt{\hat{s}}\right)+(a\leftrightarrow b).
  \label{LIdef}
\end{equation}
Here $a$ and $b$ are the two partons in partonic channel $I$;
$f_a(x,Q)$ the parton distribution function with momentum fraction $x$
and factorisation scale $Q$; and $\sqrt{\hat{s}}$ is the partonic
centre of mass energy (equal to the top quark invariant
mass $M$ at LO). Finally, $\tau$ is defined as
\begin{equation}
  \tau=\frac{\hat{s}}{s},
  \label{taudef}
\end{equation}
where $\sqrt{s}$ is the hadronic centre of mass energy of 13TeV. Note
that the definition of the luminosity in eq.~(\ref{LIdef}) already
symmetrises over the $q\bar{q}$ and $\bar{q}q$ channels, where the
prefactor removes overcounting for the $gg$ channel. From
eq.~(\ref{Rhad}), one may derive the normalised hadronic density
matrix
\begin{equation}
  \rho(M,\hat{\vec{k}})=\sum_{I\in\{q\bar{q},gg\}}
  w_I(M,\hat{\vec{k}})\rho^I(M,\hat{\vec{k}}),
  \label{rhohad}
\end{equation}
where $\rho^I(M,\hat{\vec{k}})$ is the partonic density matrix of
eq.~(\ref{rhoIdef}), and we have followed ref.~\cite{Afik:2020onf} in
defining the parton weight functions
\begin{equation}
  w_I(M,\hat{\vec{k}})=\frac{L^I(M)\tilde{A}^I(M,\hat{\vec{k}})}
  {\sum_J L^J(M)\tilde{A}^J(M,\hat{\vec{k}})}.
  \label{wIdef}
\end{equation}
In order to evaluate these expressions, we have used the publicly
available \texttt{ManeParse} package for
Mathematica~\cite{Clark:2016jgm}. Following ref.~\cite{Aoude:2022imd},
we use the \texttt{NNPDF4.0} NNLO PDF set~\cite{NNPDF:2021njg} with
strong coupling constant $\alpha_s(m_Z)=0.118$, and using the LHAPDF
interface~\cite{Buckley:2014ana}. Results for the hadronic magic,
analogous to the partonic results of figure~\ref{fig:magic}, are shown
in figure~\ref{fig:magicpp}. Interestingly, we see that the magic
still vanishes in the upper-right region of the plot, corresponding to
high top invariant mass, and forward production. Such configurations
correspond to the so-called {\it Regge limit}, in which the parton
luminosity will be dominated by gluons. Indeed, the hadron-level plot
in the upper-right region is extremely similar in structure to the
purely gluonic result of figure~\ref{fig:magic}(b).

Different behaviour is seen at low values of $\beta$ and / or $z$,
where none of the regions of low magic from figure~\ref{fig:magic}(b)
survive. Indeed, the magic no longer vanishes at $(z,\beta)=0$, for
which analytic insight may be gained as follows. Let us first define
the coefficients
\begin{equation}
  C^{\rm had.}_{ij}(z,\beta)=\frac{w_{q\bar{q}}
    (z,\beta)\tilde{C}^{q\bar{q}}_{ij}}
  {\tilde{A}^{q\bar{q}}}+\frac{w_{gg}(z,\beta)\tilde{C}^{gg}_{ij}}
  {\tilde{A}^{gg}}.
  \label{Chaddef}
\end{equation}
An explicit computation then shows that the hadron-level magic is
given by
\begin{equation}
  \tilde{M}_2^{\rm had.}(\rho)=-\log_2\left(
  \frac{1+[C_{nn}^{\rm had.}]^4+[C_{kk}^{\rm had.}]^4
    +[C_{rr}^{\rm had.}]^4+2[C_{rk}^{\rm had.}]^4}
       {1+[C_{nn}^{\rm had.}]^2+[C_{kk}^{\rm had.}]^2
    +[C_{rr}^{\rm had.}]^2+2[C_{rk}^{\rm had.}]^2}
  \right).
  \label{M2had}
\end{equation}
Using eqs.~(\ref{coeffsgg}, \ref{coeffsqq}, \ref{Fggdef}), one then
finds
\begin{equation}
  \tilde{M}_2^{\rm had}\Big|_{(z,\beta)\rightarrow(0,0)}=
  -\log_2\left(\frac{2w_{gg}^4(0,0)+w_{q\bar{q}}^4(0,0)
    +6w_{gg}^2(0,0)w_{q\bar{q}}^2(0,0)}{2w_{gg}^2(0,0)w_{q\bar{q}}^2(0,0)}
  \right).
  \label{M2hadlim}
\end{equation}
Given that both $w_{gg}(0,0)$ and $w_{q\bar{q}}(0,0)$ are non-zero,
eq.~(\ref{M2hadlim}) will also return a non-zero result. The
individual partonic cases correspond to $w_{gg}\rightarrow 0$,
$w_{q\bar{q}}\rightarrow 1$ (or vice versa), in which case one
recovers $\tilde{M}_2\rightarrow 0$ as expected. It may na\"{i}vely
seem surprising that combining the partonic channels from
figure~\ref{fig:magic}(a) and (b) does not reproduce a feature that
appears in both plots (i.e. the vanishing of the magic at
$(z,\beta)=(0,0)$). However, it is important to recognise that, whilst
the total hadronic density matrix is a simple weighted sum of the
individual partonic channels, the same is not true for the
magic. Indeed, there is a simple physical explanation for this
behaviour: eqs.~(\ref{Rq00}, \ref{rhogglims}) show that {\it
  different} stabiliser states occur in the $(z,\beta)=(0,0)$ limit
for the $gg$ and $q\bar{q}$ channels. The fact that these are
superposed with non-zero weights in the hadronic density matrix mean
that the resulting state is non-stabiliser, and thus has non-zero
magic. To see this in more detail, recall that the Pauli spectrum of a
stabiliser state has entries $\pm1$ only. Adding together {\it
  different} stabliser states with non-unit weights will therefore
result in a Pauli spectrum containing non-unit values, which is
necessarily non-stabiliser. We examine a further case of magic
enhancement upon combining states in the following section.

\subsection{Angular-averaged final states}
\label{sec:angav}

\begin{figure}
    \centering
    \subfloat[]{\includegraphics[width=0.4\textwidth]{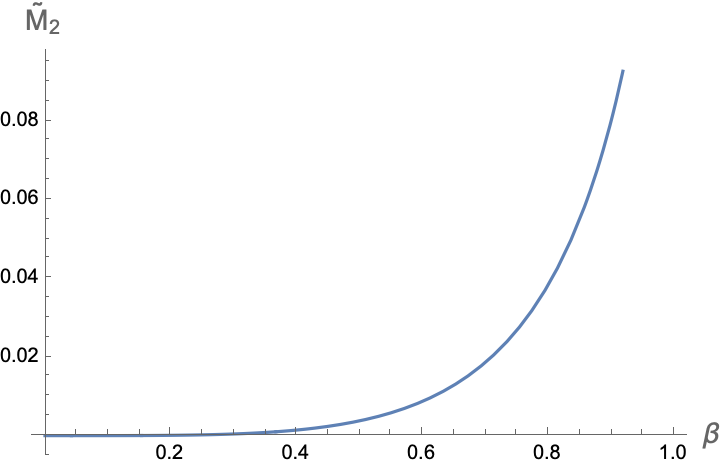} }
    \qquad
    \subfloat[]{\includegraphics[width=0.4\textwidth]{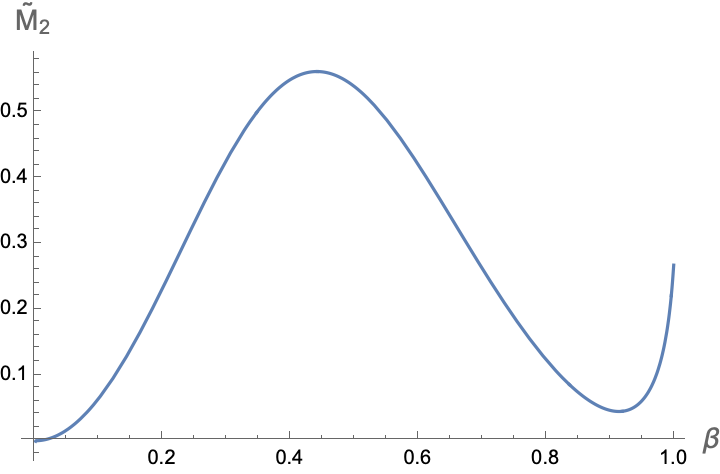} }
    \caption{The magic of an angular-averaged mixed top-antitop final
      state in: (a) the $q\bar{q}$ channel; (b) the $gg$ channel.}
    \label{fig:angavmagic}
\end{figure}

So far, we have considered the magic of the top-antitop final state
fully differentially in the kinematics (invariant mass and scattering
angle). As stated in eq.~(\ref{rhobarI}), however, one may also wish
to consider the top quark state after averaging over the top quark
directions. To do this, one may consider a fixed spin basis, rather
than the direction-dependent helicity basis. A common choice is the
{\it beam basis}, in which the $3$-direction in spin space is aligned
with the first incoming beam direction (corresponding with the
$z$-direction in Cartesian coordinates), and the $1$- and
$2$-directions (corresponding with $x$ and $y$ directions in space)
are chosen in the transverse plane so as to make a right-handed
coordinate system. The partonic angular-averaged $R$-matrices will be
given by expressions similar to eq.~(\ref{Rdecomp}), and results for
the various coefficients have been presented in
ref.~\cite{Afik:2020onf}. One finds that the $\tilde{B}^{I\pm}_i$
coefficients vanish, and that the correlation matrix
$\tilde{C}^I_{ij}$ becomes diagonal:
\begin{equation}
  \tilde{C}^I_{ij}=\left(\begin{array}{ccc}
    \tilde{C}_\perp^{I} & 0 & 0\\
    0 & \tilde{C}_\perp^{I} & 0\\
    0 & 0 & \tilde{C}_z^{I}
  \end{array}\right),
  \label{Cdiag}
\end{equation}
where we have adopted a common notation for the diagonal elements
(n.b. the elements $\tilde{C}_{xx}^I=\tilde{C}_{yy}^I\equiv
\tilde{C}_\perp^I$ are equal). The non-zero coefficients are given by
\begin{align}
  \tilde{A}^{q\bar{q}}(M)&=\frac19\left[1-\frac{\beta^2}{3}\right];\notag\\
  C_{\perp}^{q\bar{q}}(M)&=\frac{2}{135}f(\beta);\notag\\
  C_z^{q\bar{q}}(M)&=\frac{1}{9}\left[1-\frac{\beta^2}{3}
    -\frac{4}{15}f(\beta)\right];
  \label{angavqq}
\end{align}
and
\begin{align}
  \tilde{A}^{gg}(M)&= \frac{1}{192}\left[-59+31\beta^2+(66-36\beta^2+2\beta^4)
    \frac{\tanh^{-1}(\beta)}{\beta}\right];\notag\\
  C_{\perp}^{gg}(M)&=\frac{1-\beta^2}{192}\left[
    9-\frac{16\tanh^{-1}(\beta)}{\beta}
    \right]+g(\beta);\notag\\
  C_z^{gg}(M)&=\frac{1}{192}\left[-109+49\beta^2+(102-72\beta^2+2\beta^4)
    \frac{\tanh^{-1}(\beta)}{\beta}\right]-2g(\beta)
  \label{angavgg}
\end{align}
in the $q\bar{q}$ and $gg$ channels respectively, where we have
followed ref.~\cite{Afik:2020onf} in introducing the functions
\begin{align}
  f(\beta)&=\frac{(1-\sqrt{1-\beta^2})}{2};\notag\\
  g(\beta)&=\frac{f(\beta)}{96\beta^4}\left[
    49-\frac{149}{3}\beta^2+\frac{24}{5}\beta^4-(17\beta^4-66\beta^2+49)
    \frac{\tanh^{-1}(\beta)}{\beta}
    \right].
  \label{fgdef}
\end{align}
Armed with these results, we can calculate the partonic
angular-averaged magic using eq.~(\ref{M2tildedef}), which for each partonic channel $I$ yields
\begin{equation}
  \tilde{M}_2(\rho^I)=-\log_2\left(\frac{
    (\tilde{A}^I)^4+(\tilde{C}_z^I)^4+2(\tilde{C}_\perp^I)^4}
  {(\tilde{A}^I)^2[\tilde{A}^I)^2+(\tilde{C}_z^I)^2+2(\tilde{C}_\perp^I)^2]}
  \right).
  \label{M2angavresult}
\end{equation}
Results are shown in figure~\ref{fig:angavmagic}.
We see that the magic vanishes close to threshold ($\beta\rightarrow
0$) in both channels. Indeed, the analytic forms of the $R$-matrices
are found to be
\begin{equation}
  R^{q\bar{q}}(0)\propto \left(\begin{array}{cccc}1 & 0 & 0 & 0\\
    0 & 0 & 0 & 0\\
    0 & 0 & 0 & 0\\
    0 & 0 & 0 & 1
  \end{array}\right),\quad
  R^{gg}(0)\propto \left(\begin{array}{cccc}0 & 0 & 0 & 0\\
    0 & 1 & -1 & 0\\
    0 & -1 & 1 & 0\\
    0 & 0 & 0 & 0
  \end{array}\right),  
  \label{Rlimsangav}
\end{equation}
such that the quantum states can be written (using similar methods to
section~\ref{sec:vanish}) as
\begin{equation}
  R^{q\bar{q}}(0)\propto \frac{|\uparrow\uparrow\rangle
    \langle\uparrow\uparrow|
    +|\downarrow\downarrow\rangle\langle
    \downarrow\downarrow|}{\sqrt{2}},\quad
  R^{gg}(0)\propto |\tilde{\Psi}^-\rangle\langle\tilde{\Psi}^-|,\quad
  |\tilde{\Psi}^-\rangle=\frac{|\downarrow\uparrow\rangle
    -|\uparrow\downarrow\rangle}{\sqrt{2}},
  \label{Rlimsangav2}
\end{equation}
where we have used $|\uparrow\rangle$ and $|\downarrow\rangle$ to
denote single qubit spin up and down states along the beam
direction. We note that it may seem surprising from
figure~\ref{fig:magic}(a) that an angular average of the magic in the
$q\bar{q}$ channel leads to a zero result at $\beta=0$, given that
there are clearly non-zero values of the magic along the line
$\beta=0$. However, we remind the reader that we are now calculating
the magic in a different basis.

\begin{figure}
  \begin{center}
    \scalebox{0.4}{\includegraphics{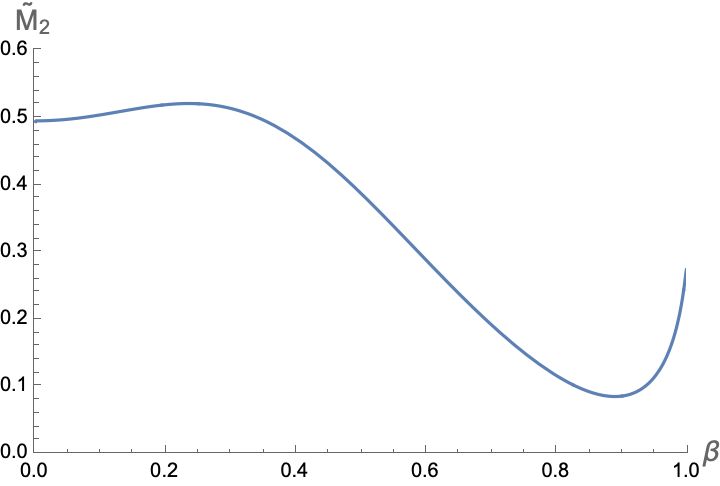}}
    \caption{The magic of a mixed angular-averaged top-antitop final
      state at hadron level.}
    \label{fig:angavmagicpp}
  \end{center}
\end{figure}

As for the fully differential magic, we may convert the partonic
magic with angular-averaging into a hadron-level result using the
appropriate analogue of eq.~(\ref{M2had}). First, one defines weighted
combinations of normalised coefficients following eq.~(\ref{Chaddef}),
starting from the coefficients of the angular-averaged density matrix:
\begin{equation}
  C_{ij}^{\rm had.}=\frac{w_{q\bar{q}}(\beta)\tilde{C}^{q\bar{q}}_{ij}}
  {\tilde{A}^{q\bar{q}}}+\frac{w_{gg}(\beta)\tilde{C}^{gg}_{ij}}
  {\tilde{A}^{gg}},
  \label{Cijnorm2}
\end{equation}
where the relevant partonic weights are now functions only of $\beta$,
due to being defined (similarly to eq.~(\ref{wIdef})) in terms of
angular-averaged quantities:
\begin{equation}
  w_I(M)=\frac{L^I(M)\tilde{A}^I(M)}{\sum_J L^J(M)\tilde{A}^J(M)}.
  \label{wIdef2}
\end{equation}
The hadronic angular-averaged magic is then given by
\begin{equation}
  \tilde{M}_2^{\rm had.}(\rho)=-\log_2\left(
  \frac{1+[C_{z}^{\rm had.}]^4+2[C_{\perp}^{\rm had.}]^4}
       {1+[C_{z}^{\rm had.}]^2+2[C_{\perp}^{\rm had.}]^2}
  \right).
  \label{M2had2}
\end{equation}
Results are shown in figure~\ref{fig:angavmagicpp}, and we note the
interesting feature that the magic is non-zero for all values of
$\beta$. This is similar to the behaviour already observed for the
fully differential magic, namely that some regions of vanishing magic
do not survive upon forming the hadronic result. The underlying
mechanism is indeed similar: superposing states from different
partonic channels leads to removal of pure stabiliser combinations, and
hence increasing magic. The results of this section are still
differential in the top quark invariant mass, and one may go further
in defining a total top quark final state by integrating over a range
of invariant masses e.g. all possible values of $\beta$, or some
suitable fiducial range. Given that angular averaging already removes
all regions of non-zero magic, however, we do not foresee any
qualitative difference upon the further mixing of states that occurs
upon performing additional integrations.

Another interesting feature of figure~\ref{fig:angavmagicpp} is that
the region of maximum magic is typically close to threshold (low
values of $\beta$), despite the fact that the magic vanishes in this
region for each partonic channel separately. As mentioned previously,
the generation of non-zero magic results from combining non-equal
stabiliser states with non-integer weights. Our results are also
consistent with ref.~\cite{Aoude:2022imd}, which shows that top quark
entanglement (as measured by the so-called {\it concurrence}) is no
longer maximal in any region once angular averaging has taken
place. Given that maximal entanglement is associated with a stabiliser
state, this again suggests that the magic will be everywhere non-zero.

\section{Conclusions}
\label{sec:discuss}
In this paper, we have performed a case study of the {\it magic}, a
property that characterises how different quantum states are from a
restricted set of so-called {\it stabiliser states}. Non-stabiliser
states are crucial for designing quantum computers that have a genuine
computational advantage over their classical counterparts, and also
for fault-tolerant algorithms. Learning how to quantify and control
magic in real-world quantum systems are both open problems, such that
it makes sense to ascertain whether magic is manifest in the quantum
playground offered by collider physics experiments.

We have focused on (anti-)top quark pair production at the LHC, which
has recently been the subject of many studies examining top quark
entanglement. Magic offers a complementary view to entanglement, in
that maximally entangled states typically have zero magic. We find
that magic top pairs are generically produced, where the amount of
magic depends on the kinematics (i.e. the scattering angle and
invariant mass). In each separate partonic channel, regions of zero
magic appear, that can be directly traced to the fact that the top
quark final state becomes a stabiliser state, which may or may not be
entangled. These regions tend to disappear upon combining channels to
form the proton-level process, or when averaging over top quark
directions. This makes sense, given that combining different states
(including individual stabiliser states) will generically result in
states which are non-stabiliser, and thus have non-zero
magic. Interestingly, the highest magic at angular-averaged level is
found close to threshold, where the magic vanishes in each separate
channel. As we have discussed in detail, this is a direct consequence
of the non-integer partonic weights multiplying each (stabiliser)
state.

There are many possible directions for future work. First, one might
look at the effect of theories beyond the Standard Model (as has been
done for entanglement in ref.~\cite{Aoude:2022imd}), in order to see
whether magic is useful as an observable that distinguishes new
physics from old. If the SSRE is insufficient, one could consider
higher order Stabiliser R\'{e}nyi entropies as additional
observables. We hope in addition that our results will prove useful in
informing a more general body of work mapping ideas from quantum
computation / information theory to collider physics, where there are
presumably many exciting interdisciplinary insights to be uncovered. 

\section*{Acknowledgments}

CDW is supported by the UK Science and Technology
Facilities Council (STFC) Consolidated Grant ST/P000754/1 ``String
theory, gauge theory and duality''. MJW is supported by the Australian Research Council grants CE200100008 and DP220100007. 

\bibliography{refs}
\end{document}